\documentclass{svproc}
\usepackage{physics}
\usepackage{epstopdf}
\usepackage{dsfont}
\usepackage{subfig}
\usepackage{graphicx}
\usepackage{amsfonts}
\usepackage{float}
\usepackage{url}

\begin{document}
\mainmatter              
\title{A new method for constructing squeezed states for the isotropic 2D harmonic oscillator}
\titlerunning{2D coherent and squeezed states}  
%
\author{James Moran\inst{1,2} \and V\'eronique Hussin\inst{1,3}}
\authorrunning{James Moran and V\'eronique Hussin} 
%
\tocauthor{Ivar Ekeland, Roger Temam, Jeffrey Dean, David Grove,
Craig Chambers, Kim B. Bruce, and Elisa Bertino}
\institute{Centre de recherches math\'ematiques, Universit\'e de Montr\'eal,  C. P. 6128, Succ. Centre-ville, Montr\'eal, Qu\'ebec, H3C 3J7, Canada,\\
\email{james.moran@umontreal.ca, hussin@dms.umontreal.ca},\\ 
\and
D\'epartement de physique, Universit\'e de Montr\'eal, C. P. 6128, Succ. Centre-ville, Montr\'eal, Qu\'ebec, H3C 3J7, Canada,
\and
D\'epartement de math\'ematiques et de statistique, Universit\'e de Montr\'eal,  C. P. 6128, Succ. Centre-ville, Montr\'eal, Qu\'ebec, H3C 3J7, Canada}



\maketitle              

\begin{abstract}
We introduce a new method for constructing squeezed states for the 2D isotropic harmonic oscillator. Based on the construction of coherent states in \cite{isoand}, we define a new set of ladder operators for the 2D system as a linear combination of the $x$ and $y$ ladder operators and construct the $SU(2)$ coherent states. The new ladder operators are used for generalizing the squeezing operator to 2D and the $SU(2)$ coherent states play the role of the Fock states in the expansion of the 2D squeezed states. We discuss some properties of the 2D squeezed states.
\end{abstract}
\section{Introduction}

Degeneracy in the spectrum of the Hamiltonian is one of the first problems we encounter when trying to define a new type of coherent states for the 2D oscillator. As a continuation of the work in \cite{isoand} we produce a non-degenerate number basis ($SU(2)$ coherent states) for the 2D isotropic harmonic oscillator with accompanying generalized creation and annihilation operators. The squeezed states for the 2D isotropic harmonic oscillator are then defined in terms of the $SU(2)$ coherent states and generalized ladder operators. 

Work on degeneracy in coherent state theory has been done, Klauder described coherent states of the hydrogen atom \cite{Klauder_1996} which preserved many of the usual properties required by coherent state analysis \cite{Ali:2012:CSW:2464906}. Fox and Choi proposed the Gaussian Klauder states \cite{PhysRevA.64.042104}, an alternative method for producing coherent states for more general systems with degenerate spectra. An analysis of the connection between the two definitions was studied in \cite{doi:10.1063/1.2435596}.

In the first part of the paper we address the degeneracy in the energy spectrum by constructing non-degenerate states, the $SU(2)$ coherent states, and we define a generalized ladder operator formed from a linear combination of the 1D ladder operators with complex coefficients.

In the last part of the paper we use a generalized squeezing operator and Fock space expansion to define squeezed states for the 2D system. In both cases we use the same definitions as for the 1D squeezed states, but replacing the Fock states with the $SU(2)$ coherent states and the 1D ladder operators with the new generalised ladder operators. We discuss the spatial probability distributions of the 2D squeezed states, as well as their dispersions.
\section{Squeezed states of the 1D harmonic oscillator}\label{1d}
Squeezed states, or squeezed coherent states, are a generalization of the standard coherent states first studied by Schr\"odinger \cite{1926NW.....14..664S}, and then formalised in the context of quantum optics by Glauber and Sudarshan \cite{PhysRev.130.2529} \cite{PhysRevLett.10.277}. In terms of the displacement and squeezing operators $D(\psi)=e^{\psi a^{\dagger} - \bar{\psi}a}$, $S(\xi)=e^{\frac{1}{2}(\xi a^{\dagger ^2} -\bar{\xi}a^2)}$ respectively, where $a, a^\dagger$ are the annihilation and creation operators, squeezed states are expressed as
\begin{equation}\label{first}
 \ket{\psi,\xi}=D(\psi)S(\xi)\ket{0},
\end{equation}
$\psi, \xi \in \mathds{C}$. Writing $\xi=r e^{i\theta}$, in terms of Fock states, $\{ \ket{n}\}$, the squeezed states are given by
\begin{equation}\label{second}
 \ket{z, \gamma}= \frac{1}{\mathcal{N}(z,\gamma)}\sum_{n=0}^{\infty}\frac{1}{\sqrt{n!}}\bigg(\frac{\gamma}{2}\bigg)^{\frac{n}{2}}H_n \Big(\frac{z}{\sqrt{2\gamma}}\Big)\ket{n},
\end{equation}
where $\frac{1}{\mathcal{N}(z,\gamma)}=\frac{1}{\sqrt{\cosh r}}e^{-\frac{\abs{z}^2}{2}}e^{\frac{\tanh r}{2} \Re (e^{i\theta}\bar{z}^2)}$. The states in equation \eqref{second} are solutions to the eigenvalue equation
\begin{equation}\label{ssevalue}
(a+\gamma {a^{\dagger}})\ket{z, \gamma}=z \ket{z, \gamma}.
\end{equation}
Equivalence between definitions \eqref{first} and $\eqref{second}$ is understood through the following relationships between the parameters \cite{gerry_knight_2004}
\begin{equation}\label{providedtransfs}
    \begin{split}
        &z=\psi-\bar{\psi}e^{i\theta}\tanh r,\\
        &\gamma=-e^{i\theta}\tanh r.
    \end{split}
\end{equation}

The term `squeezing'\ is used because the squeezed states saturate the Robertson-Schr\"odinger uncertainty relation \cite{PhysRev.34.163} but with unequal dispersions in position and momentum (unlike the standard coherent states which saturate the Heisenberg uncertainty principle with equal dispersions). The squeezed states have the following dispersions
\begin{equation}\label{1ddis}
\begin{split}
&(\Delta X)^2_{\ket{\psi,\xi}} = \bra{\psi,\xi}X^2 -\left< X\right>^2\ket{\psi,\xi}=\frac{1}{2} + \sinh ^2 r +\Re(e^{i\theta})\cosh r \sinh r ;\\
&(\Delta P)^2_{\ket{\psi,\xi}} = \bra{\psi,\xi}P^2 -\left< P\right>^2\ket{\psi,\xi}=\frac{1}{2} +\sinh ^2 r- \Re(e^{i\theta})\cosh r \sinh r,
\end{split}
\end{equation}
where $(\Delta \hat{O})^2_{\ket{\psi}} \equiv \bra{\psi}\hat{O}^2-\langle \hat{O}\rangle^2\ket{\psi}$ is the variance of the operator $\hat{O}$ in the state $\ket{\psi}$. The position and momentum operators are expressed in the usual way $\hat{X}=\frac{1}{\sqrt{2}}(a^\dagger + a)$, $\hat{P}=\frac{1}{\sqrt{2}i}(a -a^\dagger)$. When the squeezing is purely real $\xi=r$, the dispersions become $(\Delta X)^2_{\ket{\psi,\xi}}=\frac{1}{2}e^{-2r}$, $(\Delta P)^2_{\ket{\psi,\xi}}=\frac{1}{2}e^{2r}$, in this case the squeezed states saturate the Heisenberg uncertainty relation $(\Delta X)^2_{\ket{\psi,\xi}} (\Delta P)^2_{\ket{\psi,\xi}}=\frac{1}{4}$.

Like the standard coherent states, the squeezed states are also non-orthogonal and they admit a resolution of the identity \cite{cswavelets}, therefore they represent an over-complete basis for the Hilbert space of the 1D harmonic oscillator.
\section{The 2D oscillator}\label{2dosc}
For a 2D isotropic oscillator we have the quantum Hamiltonian
\begin{equation}
 \hat{H}=-\frac{1}{2}\frac{d^2}{dx^2} - \frac{1}{2}\frac{d^2}{dy^2} +\frac{1}{2} x^2 +\frac{1}{2} y^2
\end{equation}
where we have set $\hbar=1$ and the mass $m=1$ and the frequency $\omega=1$. We solve the time independent Schr\"odinger equation $H\ket{\Psi}=E\ket{\Psi}$ and obtain the usual energy eigenstates (or Fock states) labelled by $\ket{\Psi}=\ket{n,m}$ with eigenvalue $E_{n,m}=n+m+1$ and $n,m \in \mathds{Z}^{\geq0}$. These states may all be generated by the action of the raising and lowering operators in the following way \cite{Dirac1930-DIRTPO}
\begin{equation}\label{1dstuff}
\begin{split}
& a_x^-\ket{n,m}=\sqrt{n}\ket{n-1,m},\,\, a_x^+\ket{n,m}=\sqrt{n+1}\ket{n+1,m};\\ & a_y^-\ket{n,m}=\sqrt{m}\ket{n,m-1},\,\, a_y^+\ket{n,m}=\sqrt{m+1}\ket{n,m+1}.
\end{split}
\end{equation}
The states $\ket{n,m}$ in configuration space have the following wavefunction
\begin{equation}\label{theusual}
    \bra{x,y}\ket{n,m}=\psi_n(x) \psi_m(y)=\frac{1}{\sqrt{2^{n+m}n!m!}}\sqrt{\frac{1}{\pi}}e^{-\frac{ x^2}{2}-\frac{ y^2}{2}}H_n\left( x \right)H_m\left( y \right),
\end{equation}
where $\psi_n(x)= \frac{1}{\sqrt{2^{n}n!}}\left(\frac{1}{\pi}\right)^{\frac{1}{4}}e^{-\frac{ x^2}{2}}H_n\left( x \right)$ is the wavefunction of the 1D oscillator and $H_n (x)$ are the Hermite polynomials. For the physical position and momentum operators, $\hat{X}_i=\frac{1}{\sqrt{2}}(a_i^+ + a_i^-)$, $\hat{P}_i=\frac{1}{\sqrt{2}i}(a_i^- -a_i^+)$, respectively in the $i$ direction, the states $\ket{n,m}$ have the following dispersions
\begin{equation}\label{x}
    (\Delta \hat{X})^2_{\ket{n,m}}= (\Delta\hat{P}_x)^2_{\ket{n,m}}=\frac{1}{2} +n;
\end{equation}
\begin{equation}\label{y}
   (\Delta \hat{Y})^2_{\ket{n,m}}= (\Delta\hat{P}_y)^2_{\ket{n,m}}=\frac{1}{2} +m.
\end{equation}
They satisfy the Heisenberg uncertainty relation $(\Delta \hat{X})_{\ket{n,m}}(\Delta\hat{P}_x)_{\ket{n,m}}=\frac{1}{2} +n$ which grows linearly in $n$ in the $x$ direction. Similarly for the $Y$ quadratures, we obtain $(\Delta \hat{Y})_{\ket{n,m}}(\Delta\hat{P}_y)_{\ket{n,m}}=\frac{1}{2} +m$.

In what follows we will construct two new ladder operators as linear combinations of the operators in \eqref{1dstuff} and proceed to define a single indexed Fock state for the 2D system which yields the $SU(2)$ coherent states. The new ladder operators and $SU(2)$ coherent states are used to extend the definitions of the 1D squeezed states in Section (\ref{1d}) to the 2D oscillator.
\section{$SU(2)$ coherent states}
We use the ladder operators presented in Section (\ref{2dosc}) to construct a single set of creation and annihilation operators for the 2D oscillator. Introducing a set of states $\{ \ket{\nu} \}$, and defining a new set of ladder operators through their action on the set,
\begin{equation}\label{main}
   A^-\ket{\nu}=\sqrt{\nu}\ket{\nu-1},  A^+ \ket{\nu}=\sqrt{\nu+1}\ket{\nu+1}, \qquad \bra{\nu}\ket{\nu}=1, \qquad \nu=0,1,2, \ldots 
\end{equation}
These states have a linear increasing spectrum $E_\nu= \nu +1$. We may build the states by hand starting with the only non-degenerate state, the ground state, $\ket{0}\equiv \ket{0,0}$ and we take simple linear combinations of the 1D ladder operators
\begin{equation}\label{introd}
\begin{split}
    &A^+_{\alpha,\beta} = \alpha a_x^+ \otimes \mathds{I}_y +\mathds{I}_x \otimes\beta a_y^+;\\
    &A^-_{\alpha,\beta}  = \bar{\alpha} a_x^- \otimes \mathds{I}_y+\mathds{I}_x \otimes\bar{\beta} a_y^-;\\
    &[A^-_{\alpha,\beta},A^+_{\alpha,\beta}]=(\abs{\alpha}^2 + \abs{\beta}^2)\mathds{I}_x \otimes \mathds{I}_y\equiv \mathds{I},
    \end{split}
\end{equation}
for $\alpha,\beta \in \mathds{C}$, $\mathds{I}_x \otimes \mathds{I}_y=\mathds{I}_y \otimes \mathds{I}_x\equiv \mathds{I}$ and normalization condition $\abs{\alpha}^2 + \abs{\beta}^2=1$. Constructing the states $\{ \ket{\nu} \}$ starting with the ground state gives us the following table
\begin{table}[H]
\begin{center}
\begin{tabular}{ c|c }
 $\ket{\nu}$ & $\ket{n,m}$ \\
 \hline
 $\ket{0}$ & $\ket{0,0}$ \\
 $\ket{1}$ & $\alpha \ket{1,0} + \beta \ket{0,1}$ \\
 $\ket{2}$ & $\alpha^2\ket{2,0} +\sqrt{2}\alpha\beta\ket{1,1}+ \beta^2\ket{0,2}$\\
 \vdots & \vdots \\
 $\ket{\nu}$ & $\sum_{n,m}^{n+m=\nu} \alpha^n  \beta^m \sqrt{\binom{\nu}{n}} \ket{n,m}$ 
\end{tabular}%
\end{center}\caption{Construction of the states $\ket{\nu}_{\alpha,\beta} $ using the relation $A^+_{\alpha,\beta}  \ket{\nu}_{\alpha,\beta} =\sqrt{\nu+1}\ket{\nu+1}_{\alpha,\beta} $.}\label{table1}%
\end{table}%
\vspace{-2em}\noindent The states, $\ket{\nu}$, in Table \ref{table1} depend on $\alpha, \beta$ and may be expressed as
\begin{equation}
    \ket{\nu}_{\alpha,\beta}=\sum_{n=0}^{\nu} \alpha^n  \beta^{\nu-n} \sqrt{\binom{\nu}{n}} \ket{n,\nu-n}.
\end{equation}

The states $\ket{\nu}_{\alpha,\beta}$ are precisely the $SU(2)$ coherent states in the Schwinger boson representation \cite{Ali:2012:CSW:2464906}. This makes sense from our construction, the degeneracy present in the spectrum $E_{n,m}$ is an $SU(2)$ degeneracy, and so we created states which averaged out the degenerate contributions to a given $\nu$.
These states have the following orthogonality relations
\begin{equation}
 \bra{\mu}_{\gamma,\delta}\ket{\nu}_{\alpha,\beta}=(\bar{\gamma}\alpha +\bar{\delta}\beta)^\nu \delta_{\mu,\nu},
\end{equation}
which reduces to a more familiar relation when $\gamma=\alpha$ and $\delta=\beta$
\begin{equation}
  \bra{\mu}_{\alpha,\beta}\ket{\nu}_{\alpha,\beta}=\delta_{\mu,\nu},
\end{equation}
using the normalization condition $\abs{\alpha}^2 + \abs{\beta}^2=1$. 
\begin{figure}[H]
    \centering
    \begin{minipage}{.5\textwidth}
    \centering
    \includegraphics[scale=0.5]{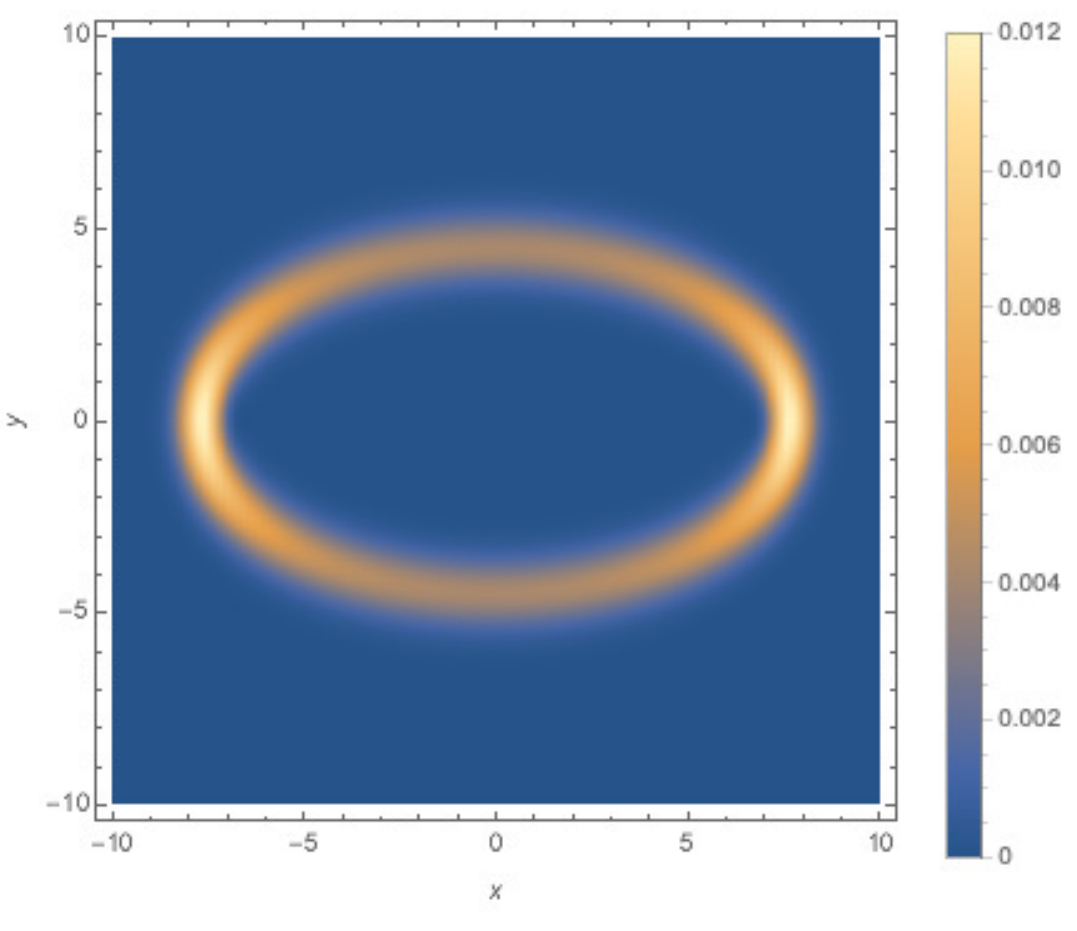}
    \end{minipage}%
    \begin{minipage}{.5\textwidth}
    \centering
    \includegraphics[scale=0.5]{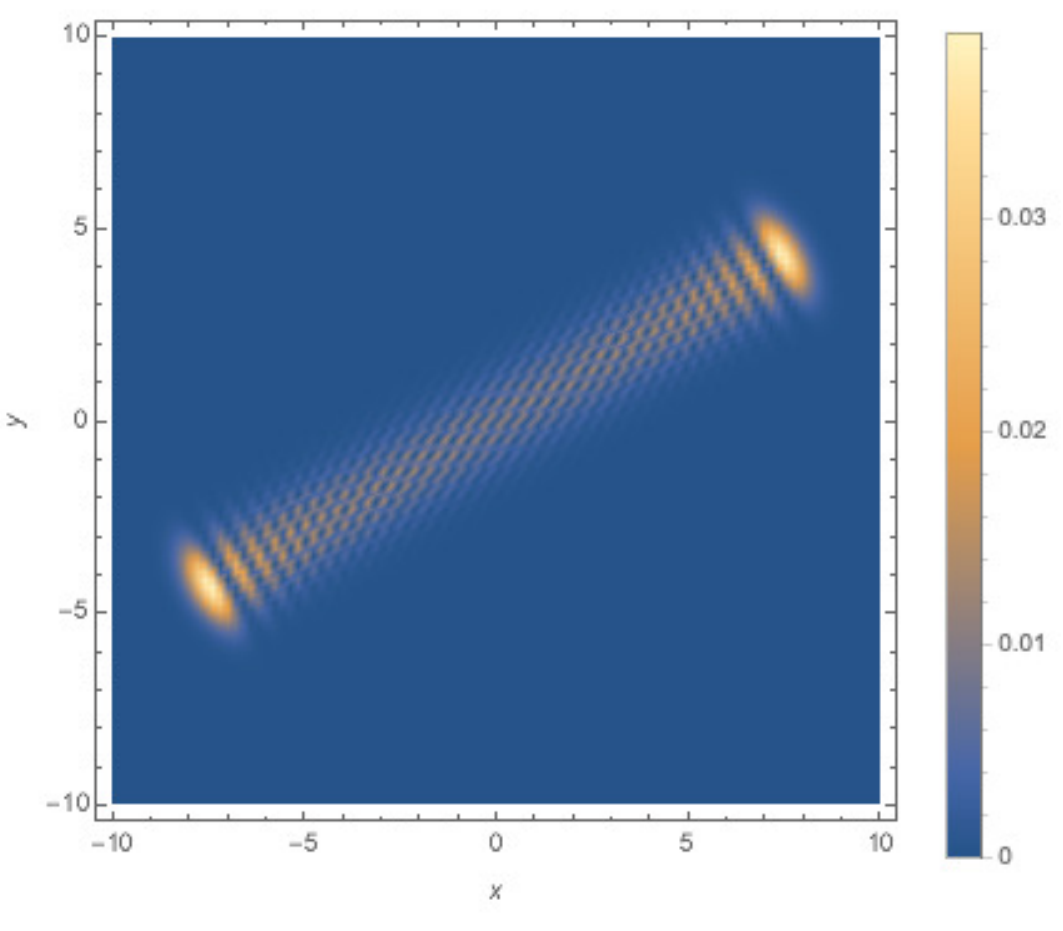}
    \end{minipage}%
    \caption{Density plots of $\abs{\bra{x,y}\ket{\nu}_{\alpha,\beta}}^2$ for $\alpha= \frac{\sqrt{3}}{2}e^{i\frac{\pi}{2}}, \beta=\frac{1}{2}$ (left) and $\alpha= \frac{\sqrt{3}}{2}, \beta=\frac{1}{2}$ (right) both at $\nu=40$.}%
    \label{fig:111}%
\end{figure}
The probability densities,  $\abs{\bra{x,y}\ket{\nu}_{\alpha,\beta}}^2$, of the quantum $SU(2)$ coherent states form ellipses when viewed as density plots, this mimics the classical 2D oscillator spatial distribution. This has been studied extensively by Chen \cite{Chen_2003}.

The $SU(2)$ coherent states have the following variances for the physical position and momentum operators $\hat{X}_i=\frac{1}{\sqrt{2}}(a_i^+ + a_i^-)$, $\hat{P}_i=\frac{1}{\sqrt{2}i}(a_i^- -a_i^+)$, respectively in the $i$ direction

\begin{equation}
    (\Delta \hat{X})^2_{\ket{\nu}_{\alpha,\beta}}= (\Delta\hat{P}_x)^2_{\ket{\nu}_{\alpha,\beta}}=\frac{1}{2} +\abs{\alpha}^2  \nu;
\end{equation}

\begin{equation}
   (\Delta \hat{Y})^2_{\ket{\nu}_{\alpha,\beta}}= (\Delta\hat{P}_y)^2_{\ket{\nu}_{\alpha,\beta}}=\frac{1}{2} +\abs{\beta}^2  \nu. 
\end{equation}
The results are essentially the same as those in \eqref{x} and \eqref{y}, but they are tuned by the continuous parameters $\alpha, \beta$ introduced in \eqref{introd}.

\section{2D squeezed states}
By analogy with the 1D case we define a 2D displacement and 2D squeezing operators
\begin{equation}
 D(\Psi)=e^{\Psi A^+_{\alpha,\beta} -\bar{\Psi}A^-_{\alpha,\beta}},
\end{equation}
and
\begin{equation}
 S(\Xi)= \exp\left( \frac{1}{2}[\Xi {A^+_{\alpha,\beta}}^2 - \bar{\Xi}{A^-_{\alpha,\beta}}^2]\right)
\end{equation}
respectively. The generalized squeezed state is obtained through the action of the two operators on the 2D vacuum
\begin{equation}
 \ket{\Psi,\Xi}_{\alpha,\beta}=D(\Psi)S(\Xi)\ket{0}_{\alpha,\beta}.
\end{equation}

Using the expansion of the 1D squeezed states, we replace the basis $\ket{n} \rightarrow \ket{\nu}_{\alpha,\beta}$ and use capital lettered parameters (to indicate they are 2D states) to get the following
\begin{equation}\label{expans}
  \ket{Z, \Gamma}_{\alpha,\beta}= \frac{1}{\sqrt{\cosh R}}e^{-\frac{\abs{Z}^2}{2}}e^{\frac{\tanh R}{2} \Re (e^{i\Theta}\bar{Z}^2)}\sum_{\nu=0}^{\infty}\frac{1}{\sqrt{\nu !}}\bigg(\frac{\Gamma}{2}\bigg)^{\frac{\nu}{2}}H_\nu \Big(\frac{Z}{\sqrt{2\Gamma}}\Big)\ket{\nu}_{\alpha,\beta},
\end{equation}
with $Z=\Psi -\bar{\Psi}e^{i\Theta}\tanh R, \Gamma=-e^{i\Theta}\tanh R$.
\begin{figure}[H]
    \centering
    \begin{minipage}{.5\textwidth}
    \centering
    \includegraphics[scale=0.5]{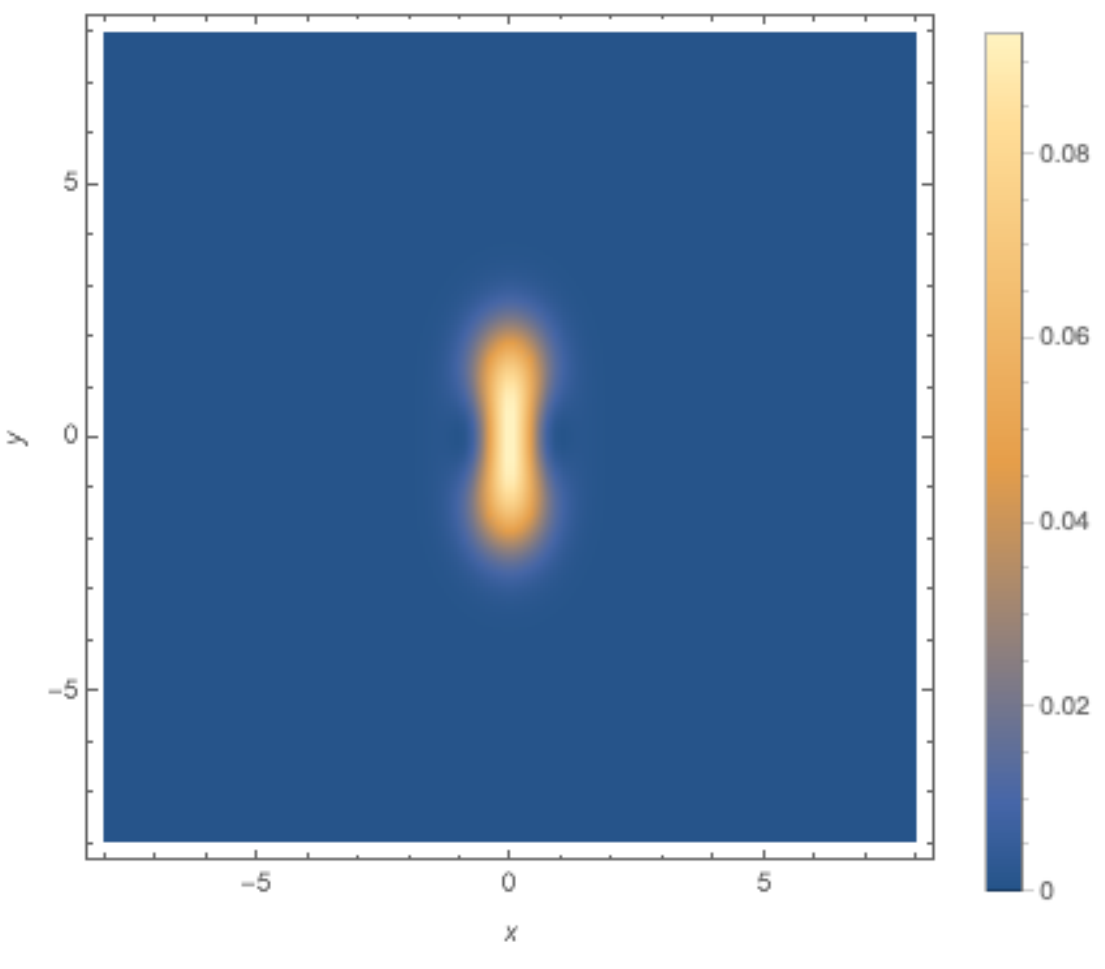}
    \end{minipage}%
    \begin{minipage}{.5\textwidth}
    \centering
    \includegraphics[scale=0.5]{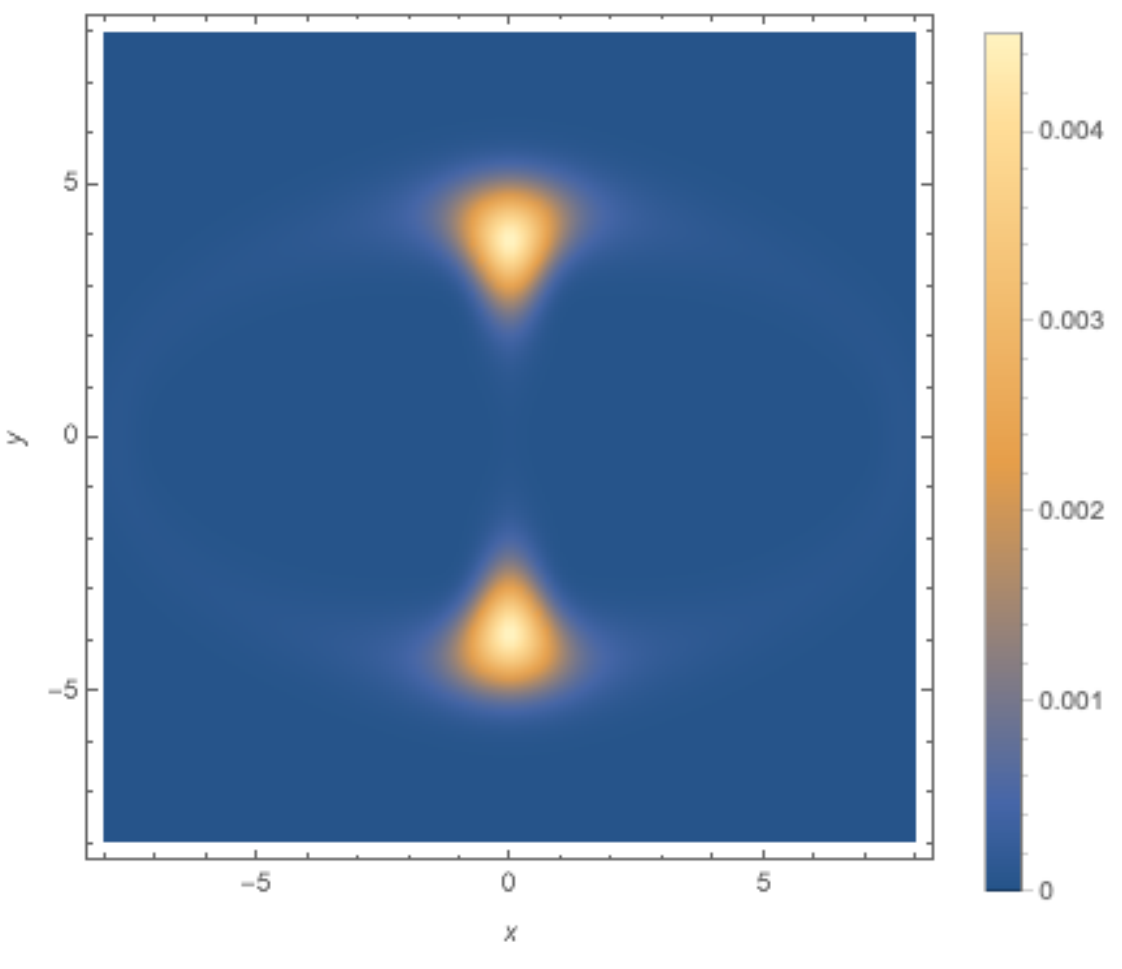}
    \end{minipage}%
    \caption{Density plots of $\abs{\bra{x,y}\ket{\Psi,\Xi}_{\alpha,\beta}}^2$ for $\alpha= \frac{\sqrt{3}}{2}e^{i\frac{\pi}{2}}, \beta=\frac{1}{2}, \Psi=1, R=0.1, \Theta=0$ (left) and $\alpha= \frac{\sqrt{3}}{2}, \beta=\frac{1}{2}, \Psi=1,R=10, \Theta=0$ (right) both with $20$ terms kept in the expansion of equation \eqref{expans}}%
    \label{fig:12}%
\end{figure}
In Figure \ref{fig:12} we see the effect of increasing the strength of the squeezing, on the left most plot the squeezing is relatively small, $R=0.1$ and the probability density is converging to a single maximum. This is in agreement with the limit $R\rightarrow0$ which would produce a Gaussian distribution with single maximum \cite{isoand}. On the other hand, the rightmost plot, $R=10$, reveals a separation of the probability density onto two distinct maxima. It is important to note that the graphs are not properly normalized as a truncated sum ($20$ terms) was used in the computation.

Restricting to the case of the 2D squeezed vacuum, $\Psi=0$, the squeezing operator admits an $su(1,1)$ decomposition \cite{PhysRevD.29.1107} yielding
\begin{equation}\label{2dsqueeze}
\ket{\Xi}_{\alpha,\beta}=\frac{1}{\sqrt{\cosh R}} \exp{\frac{e^{i\Theta}}{2}\tanh R (\alpha^2 {a_x^+}^2 + \beta^2 {a_y^+}^2+\alpha\beta a_x^+ a_y^+)} \ket{0,0}
\end{equation}
in terms of the 1D ladder operators. Equation \eqref{2dsqueeze} does not factorise, $\ket{\Xi} \not = \ket{\xi_x}_x \otimes \ket{\xi_y}_y$; the bilinear 1D terms in the expansion of $ {A^+_{\alpha,\beta}}^2$ have induced a coupling between the $x$ and $y$ modes of the oscillator. This represents a non-trivial generalization of the squeezed states to 2D, a two-mode-like squeezing was generated as a result of the construction, but the 2D squeezed states themselves retain most of the definitions of their 1D counterparts.

To calculate the dispersions in $x$ and $y$ we use the Baker-Campbell-Haussdorf identity $e^A B e^{-A} = B+[A,B] +\frac{1}{2}[A,[A,B]]+ \ldots$ \cite{griffiths_schroeter_2018} to compute Bogoliubov transformations, for example, the $x$ ladder operators are transformed as
\begin{equation}
\begin{split}
 S^\dagger (\Xi) a_x^- S(\Xi)=(&\abs{\beta}^2+\abs{\alpha}^2\cosh R) a_x^- +\alpha \bar{\beta}(\cosh R -1)a_y^-\\
 & +e^{i\Theta} \sinh R(\alpha^2 a_x^+ + \alpha\beta a_y^+);
 \end{split}
\end{equation}
\begin{equation}
\begin{split}
 S^\dagger (\Xi) a_x^+ S(\Xi)=(&\abs{\beta}^2+\abs{\alpha}^2\cosh R) a_x^+ +\bar{\alpha} \beta(\cosh R -1)a_y^+\\
 & +e^{-i\Theta} \sinh R(\bar{\alpha}^2 a_x^- + \bar{\alpha}{\beta} a_y^-).
\end{split}
\end{equation}
Using these transformations we can compute the dispersions in $x$
\begin{equation}
\begin{split}
  &(\Delta \hat{X})^2_{\ket{\Xi}_{\alpha,\beta}}=\frac{1}{2}+\abs{\alpha}^2 \sinh^2 R + \Re(e^{i\Theta} \alpha^2) \sinh R \cosh R;\\
  &(\Delta \hat{P}_x)^2_{\ket{\Xi}_{\alpha,\beta}}=\frac{1}{2}+\abs{\alpha}^2 \sinh^2 R - \Re(e^{i\Theta} \alpha^2) \sinh R \cosh R,
  \end{split}
\end{equation}
and similarly for $y$
\begin{equation}
\begin{split}
  &(\Delta \hat{Y})^2_{\ket{\Xi}_{\alpha,\beta}}=\frac{1}{2}+\abs{\beta}^2 \sinh^2 R + \Re(e^{i\Theta} \beta^2) \sinh R \cosh R;\\
  &(\Delta \hat{P}_y)^2_{\ket{\Xi}_{\alpha,\beta}}=\frac{1}{2}+\abs{\beta}^2 \sinh^2 R - \Re(e^{i\Theta} \beta^2) \sinh R \cosh R.
  \end{split}
\end{equation}
These results also hold for the generalized squeezed states $\ket{\Psi,\Xi}_{\alpha,\beta}$ because the action of the displacement operator has no effect on the on the variances. The results resemble those in equation \eqref{1ddis} but are modified by $\alpha, \beta$. We see in the limit $R\rightarrow 0$ we saturate the Heisenberg uncertainty relation in both $x$ and $y$.
\section{Conclusion}
In this paper we have described a method for constructing squeezed states for the 2D isotropic oscillator which relies on using the minimal set of definitions used to describe the squeezed states of the 1D oscillator. Unlike the coherent states defined in a similar manner in \cite{isoand}, the generalized squeezed states did not factorise into the product of squeezed states on $x$ and $y$ independently. A coupling was induced which took the form of a two-mode like squeezing creating an entanglement between the two modes.

We found the dispersions for the 2D squeezed states and saw that they resemble the dispersions in the 1D case but modified by the parameters $\alpha, \beta$ introduced during the construction of the $SU(2)$ coherent states. As well we saw a separation of the spatial probability densities into two distinct maxima for larger values of the squeezing $R$.

Finally, perhaps this method can be used to construct squeezed states for more general degenerate and higher dimensional systems and oscillators. The approach presented in this paper will require modification on a case by case basis because in general a multidimensional system will admit a more complex degenerate structure, which would significantly modify the generalzsed ladder operators as well as the non-degenerate basis $\{\ket{\nu}\}$. If a system possesses non-algebraic degeneracies, such as the 2D particle in a box (e.g. $1^2 +7^2=5^2 + 5^2$), a new method for counting states contributing to a degenerate subgroup $\ket{\nu}$ would be required.

%
\bibliographystyle{unsrt}
\bibliography{refs.bib}

\end{document}